\documentclass{article}

\usepackage[utf8]{inputenc}
\usepackage{amsthm,amsmath,amssymb}
\usepackage{graphicx}
\usepackage{hyperref}
\usepackage{authblk,setspace}
\usepackage[shortlabels]{enumitem}
\usepackage[a4paper, total={6in, 8in}]{geometry}
\usepackage[
  backend=biber,
  style=ieee,
  citestyle=numeric-comp,
  maxbibnames=4,
  sorting=none,
  sortcites=true
]{biblatex}
\usepackage{csquotes}
\usepackage{xcolor}
\usepackage{booktabs}
\usepackage{caption}
\title{Evaluating Efficiency of Platform Trials Under Delayed Outcomes Using Treatment Throughput}

\author[1]{Aritra Mukherjee}
\author[1]{James M. S. Wason}

\affil[1]{Population Health Sciences Institute, Newcastle University}

\date{}

\begin{document}

\onehalfspacing
\maketitle

\begin{abstract}
\noindent \textbf{Background}: Over the last few years, the use of platform trial designs in practice has been gaining increased momentum. This trial design allows early stopping of less promising arms, facilitates use of a shared control arm across different arms and different stages and adding of new treatment arms to an ongoing trial without harming the statistical properties of the trial. Platform trials lead to improved efficiency, patient-benefit, and reduce the time to finding one or more effective treatments in a disease area. However, like other adaptive designs these advantages may be affected adversely by a delay in observing the primary endpoint used to make interim decisions. In presence of such delay, the choice is between (a) pausing recruitment until requisite data is accrued for the interim analysis, leading to longer time to find or identify one or more effective treatment arms; or (b) continuing to recruit patients, which may result in a large number of participants who do not benefit from the interim analysis. Limited research has investigated to what extent outcome delay affects the efficiency of platform trials. Our study assesses the impact of outcome delay on these trials through a novel metric measuring the intensity of intervention arms evaluated. 

\noindent \textbf{Methods}: We propose Expected Throughput $(ET)$ as a measure of efficiency for platform trials. This is computed based on the expected number of treatment arms evaluated per 1000 participants. The metric accounts for pipeline participants accumulated whilst treatment outcome is awaited. We consider a multi-stage sequential design for each treatment arm in the study and assess $ET$ for different delay lengths. We assume uniform accrual of patients in the trial.

\noindent \textbf{Results}: The results confirm that there is a severe impact of delayed outcome on the efficiency of platform trials. The efficiency loss is highly dependent on the recruitment rate and the outcome delay length. For a moderate recruitment rate of 10 per month we see severe losses over a delay length of greater than 12 months. In general, it can be inferred that if the delay length is more than $2/3^{rd}$ of the time to recruit for a single arm, a platform trial is able to evaluate significantly lower number of treatment arms on average than initially anticipated.

\noindent \textbf{Conclusions}: The study analyses the impact of delay on platform trial through a novel and practical metric. A thorough inspection of this impact must be studied at the planning stage considering different recruitment rates to fully understand the implications of a delayed outcome. \\


\end{abstract}

\section{Introduction}

Platform trials represent an emerging class of clinical trial designs in which several interventions can be tested at the same time against a shared control group under a single overarching master protocol.
These trials are built on adaptive design principles and are often an extension of multi-arm multi-stage designs \cite{Angus2019,Choodari-Oskooei2020,Koenig2024,Lee2020,Lee2021,Meyer2021,Park2022,Roustit2023,Saville2016,Ventz2018,Woodcock2017}.
A distinguishing feature of platform trials is their added flexibility, including the ability to introduce new experimental arms and modify the control arm while the trial is ongoing.

Compared with conventional clinical trial designs platform trials provide distinct advantages. 
Rather than being primarily intervention-focused, platform trials are more disease-focused, as they support the efficient and ongoing assessment of multiple treatments and allow findings to be adapted in response to both internal trial results and external scientific developments. 
Further, platform trials enable the continuous evaluation of multiple interventions using shared infrastructure and harmonized procedures under one master protocol, reducing the administrative time required to set up new trials each time a new treatment is available.
Given the flexibility and efficiency they offer, in recent years platform trials have been gaining popularity in academic as well as industrial trials \cite{Pitre2023}. 

However, one key assumption that is often made in planning a platform trial is the primary treatment outcome is immediately available after recruiting a trial participant. 
This is typically not a valid assumption as the primary outcome might take a long time to observe.
It has been noted in recent developments \cite{Wason2019,Mukherjee2022Neuro,Mukherjee2022Cancer,Mukherjee2025GSD,Mukherjee2025MAMS} that having a delay in measuring the primary outcome data can be particularly harmful in terms of the expected efficiency gain of adaptive designs.
In presence of outcome delay, accrued pipeline participants while awaiting treatment outcomes during the interim analyses, inflate the effective sample size for the trial considerably.
That in turn either increases the ESS of the trial or the trial ends up being unnecessarily large (for sample size re-estimation designs).
Platform trials, built on these adaptive design principles can be suspected to be subject to similar efficiency losses. 
However, the earlier findings cannot readily be applied to platform trials as platform trials are not assessed based on expected time to complete the trial or ESS.
This study therefore aims to assess the impact of delay on platform trials assuming a uniform recruitment rate.

\section{Examples of platform trials with a delayed outcome}

Delayed endpoints are a common feature of platform trials, as discussed in the section above. 
Table \ref{tab: Example_platform trials} provides a few examples of MAMS platform trials across different disease areas in which the primary outcome is subject to such a delay. 
As illustrated by these examples, platform trials frequently rely on primary outcomes that may take considerable time to ascertain.
This delay length can vary between a few weeks time to years.
Consequently, even when a platform trial allows efficient recruitment and comparison of multiple interventions, the availability of usable primary outcome data and therefore the ability to make definitive treatment comparisons may remain constrained by the outcome data available to analyse.

\begin{table}[]
\centering
\caption{Examples of platform trials with a delayed outcome}
\label{tab: Example_platform trials}
\resizebox{\textwidth}{!}{%
\begin{tabular}{llllll}
\hline
\textbf{NCT Number} & \textbf{Study Title} & \textbf{Conditions} & \textbf{Primary Outcome Measures} & \textbf{\begin{tabular}[c]{@{}l@{}}Outcome \\ type\end{tabular}} & \textbf{\begin{tabular}[c]{@{}l@{}}Delay \\ length\end{tabular}} \\ \hline \\
NCT04297683 & \begin{tabular}[c]{@{}l@{}}HEALEY ALS Platform Trial - \\ Master Protocol\end{tabular} & \begin{tabular}[c]{@{}l@{}}Amyotrophic \\ Lateral\\ Sclerosis\end{tabular} & \begin{tabular}[c]{@{}l@{}}Change in disease severity as measured by\\ the ALS Functional Rating Scale-Revised \\ (ALSFRS-R) total score and survival., \\ 36 Weeks\end{tabular} & \begin{tabular}[c]{@{}l@{}}Continuous\\ (repeated \\ measures)\end{tabular} & \begin{tabular}[c]{@{}l@{}}36 weeks \\ (every \\ 4 weeks)\end{tabular} \\ \\
NCT07173803 & \begin{tabular}[c]{@{}l@{}}The Progressive Supranuclear   \\ Palsy Clinical Trial Platform\end{tabular} & \begin{tabular}[c]{@{}l@{}}Progressive \\ Supranuclear \\ Palsy(PSP)\end{tabular} & \begin{tabular}[c]{@{}l@{}}Change in disease severity as measured by\\  the 15-item modified Progressive\\ SupranuclearPalsy Rating Scale \\ (mPSPRS-15) in which the minimum \\ score is 0 and the maximum score is 52,\\ with higher scores indicating a worse \\ outcome., 52 weeks\end{tabular} & Continuous & 52 weeks \\ \\
NCT04805125 & \begin{tabular}[c]{@{}l@{}}Immunocompromised \\ Swiss Cohorts Based Trial \\ Platform\end{tabular} & \begin{tabular}[c]{@{}l@{}}Immuno-\\ compromised\\ Patients\end{tabular} & \begin{tabular}[c]{@{}l@{}}Change in pan-Ig antibody response \\ (pan-Ig anti-S1-RBD) from \\ baseline (day of vaccination) \\ to three months after \\ vaccination, measured using the \\ Elecsys® Anti-SARS-CoV-2 S \\ immunoassay.\end{tabular} & Continuous & 3 months \\ \\
NCT06461429 & \begin{tabular}[c]{@{}l@{}}PLatform for Adaptive Trials \\ In Perinatal UnitS - \\ (Core Protocol)\end{tabular} & Preterm Birth & \begin{tabular}[c]{@{}l@{}}Number of participants who progress \\ by at least one level higher on \\ the PLATIPUS Ordinal Outcome Scale\end{tabular} & Continuous & 42 weeks \\ \\
NCT05553808 & \begin{tabular}[c]{@{}l@{}}Platform Trial of  Novel Regimens \\ Versus Standard of Care (SoC) in \\ Participants With Non-small Cell \\ Lung Cancer (NSCLC)\end{tabular} & Neoplasms & \begin{tabular}[c]{@{}l@{}}Overall survival, calculated as \\ time from randomization to death. \\ Up to 2 years\end{tabular} & Time to event & 2 years \\ \hline
\end{tabular}%
}
\end{table}

Despite the potential for delayed outcomes to affect trial efficiency, the design and planning of these trials do not necessarily explicitly account for the delay in outcome availability. 
It is therefore important to understand the extent to which delayed outcomes may affect the efficiency and operational characteristics of platform trials. 
The following sections investigate this issue in detail, examining the impact of delayed outcome on key aspects of platform trial design efficiency.

\section{Methods}

\subsection{Design and notation}

Let us assume that the master protocol for the platform trial allows each individual treatment to have a maximum of $J$-stages with equally spaced interim analyses.
This implies that the $j^{th}$ interim analysis for any treatment arm would be based on $j/J$ proportion of the maximum sample size required for that arm for $j=1, \cdots,J$.
We assume there is an infinite number of new treatments available to test and  that an arm that is dropped or reaches a final analysis is immediately replaced by a new experimental treatment arm.

Let, $X_{ljk}$ denote the treatment outcome for the $l^{th}$ patient in the $k^{th}$ treatment arm at stage $j,  l=1, 2,\cdots, n_{jk}$; $j=1,2,\ldots, J; k= 1, 2, \cdots$ where $n_{jk}$ denotes the sample size for the $k^{th}$ arm at stage $j$.
Without the loss of generality, for this study, we assume equally spaced interim analyses with equal allocation ratio for control and all the experimental treatment arms, i.e. $n_{jk}=n; \quad \forall j,k$.
Further, let  $\mu_k, k=1,2,\ldots$ be the mean response on the experimental treatments and $\mu_0$ be the mean response on the control, i.e.,
\begin{equation*}
    X_{ljk} \sim N\left(\mu_k,\sigma_k^2\right);k=0,1,2,\ldots .
\end{equation*}
Then, the hypotheses test for each treatment arm are given by, $H_{0k}:\delta_k=\mu_k-\mu_0=0;k=1,2,\ldots $ vs. $H_{1k}:\delta_k=\mu_k-\mu_0> 0 ;k=1,2,\ldots $, where each treatment arm is powered $100*(1-\beta)\%$ at $\delta^*$, $\delta^*$ being the MCID.
We reject the null hypothesis for a treatment if the observed test statistic $Z_{ijk}$ [defined later in this section] exceeds the corresponding efficacy boundary at any stage.

Note that, in practice, it is not possible to have an infinite supply of trial participants at any given time point. 
Therefore, we assume a maximum of $k^*$ arms can be evaluated at any interim analyses.
This can be looked at as $k^*$ available slots at any interim analysis to test treatments.
Each treatment arm is assessed in each interim based on different pre-specified stopping rules.
We assume that the design under consideration is a MAMS platform with similar stopping boundaries for each treatment arm. 
For example, all the treatments entering the platform will be assessed based on a $J$-staged MAMS for $k^*$ treatment arms with OBF stopping bounds.
Thus, here the type I error rate is controlled at each individual treatment arm with marginally controlled power.

Now, a platform trial can be thought of as a perpetually running trial. 
However, for the ease of defining the trial design and investigate the possibilities, we restrict ourselves to looking at a particular time interval of $[0,t]$.
Based on the value of $t$, the window can be increased/ decreased as per requirement.
Let, the no. of interim analysis points between the time interval [0,t] be denoted by $J^*$.
Given, at every stage we can complete evaluation for a maximum of $k^*$ treatment arms under inspection, then, the maximum no. of treatment arms that can be evaluated in this time window is $J^* k^*$.
This would imply that every treatment stops early at its first interim analysis.

We define three vectors as follows:
\begin{itemize}
    \item $J_0=\{j_{0k} : j_{0k}$  denote the stage at which treatment $k$ was introduced in the platform trial, $j_{0k}=1,2,…,J^*  ;k=1,2,…,J^* k^*\}$
    \item $\Omega=\{\omega_k: \omega_k$  denote whether the $k^{th}$ treatment is effective or not; $k=1,2,…,J^* k^*\}$. In other words, i.e. stopped for efficacy or reached final analysis and was significant.
    \item $\Psi=\{\psi_k: \psi_k$ denotes at which stage a decision was made regarding treatment  $k$, i.e. the stage $k^{th}$ treatment was found to be effective/ineffective, i.e., $\psi_k=1,2,,…,J \text{ for } k=1,2,…,J^* k^*\}$
\end{itemize}
		
Given the values of these vectors, we can describe a platform trial for a given time interval.

Let, $Z_{ijk}$ denote the Wald’s statistic for treatment $k$, at stage $j;j=1,2, \cdots, \psi_k$, being introduced at stage $i;i=1,2, \cdots, J^*$ in the trial.
Now, $Z_{ijk}$ can be obtained as:

\begin{equation*}
 Z_{ijk}= \frac{\frac{1}{n_{jk}}\sum_{l=1}^{n_{jk}}X_{ljk}-\frac{1}{n_{0k}}\sum_{l=1}^{n_{0k}}X_{lj0}}{\sqrt{\left(\frac{\sigma_k^2}{n_{jk}}+\frac{\sigma_0^2}{n_{j0}}\right)}} ={\hat{\delta}}_{jk}{I}_{jk}^{1/2}   
\end{equation*}
   
where, ${\hat{\delta}}_{jk}=\left(\frac{1}{n_{jk}}\sum_{l=1}^{n_{jk}}X_{ljk}-\frac{1}{n_{0k}}\sum_{l=1}^{n_{0k}}X_{lj0}\right)$ is the observed treatment effect and $I_{jk}=\left(\sqrt{\left(\frac{\sigma_k^2}{n_{jk}}+\frac{\sigma_0^2}{n_{j0}}\right)}\right)^{-1}$ is the information for treatment $k$ in stage $j$.

Since we assume equally spaced designs with equal allocation across arms, $n_{jk}=j*n$ and $n_{j0}=(i+j)*n$. 
For this study, we assume analysis using concurrent controls only, meaning that $i=0$ in $n_{j0}=(i+j)*n$.

Now for any given set of $\{(j_{01},\omega_1,\psi_1 ),…,(j_{0k},\omega_k,\psi_k )\}$, the vector of test statistics $Z=(Z_{j_{01} 11},Z_{j_{01} 21},…,Z_{j_{01} \psi_1 1},Z_{j_{02} 12},Z_{j_{02} 22},…,Z_{j_{02} \psi_2 2},…,Z_{j_{0k} 1k},Z_{j_{0k} 2k},…,Z_{j_{0k} \psi_k k})'$ follows a multivariate normal distribution of dimension $\sum_1^k\psi_k$  with mean vector \[
\left(
\underbrace{\tau_1,\tau_1,\ldots,\tau_1}_{\psi_1\ \text{times}},
\underbrace{\tau_2,\tau_2,\ldots,\tau_2}_{\psi_2\ \text{times}},
\ldots,
\underbrace{\tau_k,\tau_k,\ldots,\tau_k}_{\psi_k\ \text{times}}
\right)'
\] 
and dispersion matrix $\Sigma=E(ZZ')$.

Now, the elements of the dispersion matrix can be computed as :

\[
\operatorname{Cov}\!\left(Z_{i_1j_1k_1},Z_{i_2j_2k_2}\right)=
\begin{cases}
\displaystyle
\frac{\sigma\sqrt{\dfrac{1}{n_{j_1k}}+\dfrac{1}{n_{j_10}}}}
{\sigma\sqrt{\dfrac{1}{n_{j_2k}}+\dfrac{1}{n_{j_20}}}} = \sqrt{\frac{j_2}{j_1}},
& k_1=k_2,\ i_1=i_2,\ j_1>j_2,\\[3ex]

\displaystyle
\frac{\sigma\sqrt{\dfrac{1}{n_{j_2k}}+\dfrac{1}{n_{j_20}}}}
{\sigma\sqrt{\dfrac{1}{n_{j_1k}}+\dfrac{1}{n_{j_10}}}}= \sqrt{\frac{j_1}{j_2}},
& k_1=k_2,\ i_1=i_2,\ j_1<j_2,\\[3ex]

\displaystyle
\left(I_{j_1k_1}I_{j_2k_2}\right)^{1/2}
\sigma\sqrt{\frac{1}{j_1n}},
& k_1\neq k_2,\ i_1=i_2,\ j_1>j_2,\\[3ex]

\displaystyle
\left(I_{j_1k_1}I_{j_2k_2}\right)^{1/2}
\sigma\sqrt{\frac{1}{j_2n}},
& k_1\neq k_2,\ i_1=i_2,\ j_1<j_2,\\[3ex]

\displaystyle
\left(I_{j_1k_1}I_{j_2k_2}\right)^{1/2}
\sigma\sqrt{\frac{1}{j_2n}},
& k_1\neq k_2,\ i_1>i_2,\ j_1<j_2,\ i_1+j_1=i_2+j_2,\\[3ex]

\displaystyle
\left(I_{j_1k_1}I_{j_2k_2}\right)^{1/2}
\sigma\sqrt{\frac{1}{j_1n}},
& k_1\neq k_2,\ i_1<i_2,\ j_1>j_2,\ i_1+j_1=i_2+j_2,\\[3ex]

0,
& \text{otherwise.}
\end{cases}
\]

Here, the first two cases in the above equation denotes the variance and covariances of the test statistic of the same treatment for different stages.
The next two cases denotes the the variance and covariances of the test statistic of two different treatments that start at the same stage (say $i_1=i_2=j_0$) across different stages.
The following two cases captures the variance and covariances of test statistics of two different treatments across different stages that start at different stages but have one or more overlapping stages.
The last case denotes independent test statistics which will be the case for different treatments with no overlapping stages.
Note that, this is zero only because we use concurrent controls in this study.
Therefore, it can be stated that, the test statistic for any two treatment arms:
\begin{itemize}
    \item will be independent if they enter the trial at different time points with no overlapping stages between them.
    \item will be dependent through the shared control and it will depend on the $j_1$ and $j_2^{nd}$ stage they both are in.
\end{itemize}

\subsection{Measuring efficiency in platform trial}

A platform trial can be defined as a perpetually running trial.
Since, we cannot pre-specify the duration to complete the trial or the number of treatment arms beforehand, the expected sample size or the expected duration would not be a suitable metric to compare or evaluate efficiency of multiple platform trial designs against each other. 
Since in terms of efficiency a platform trial allows to assessing multiple treatment arms under one single master protocol, it can be argued that, greater the number of treatment arms can be evaluated during a time frame, the better the platform trial is (assuming the statistical properties are the same). 
We can define the efficiency of a platform trial in terms of the expected number of treatment arms $[E_{J^*} [K_{eval})$, detailed in the following paragraphs] that can be evaluated during this time period of conducting $J^*$ analyses.

Let us assume that patient is accrued in the trial at the rate of $\lambda$ per unit of time during the time interval [0,t]. Therefore, the maximum sample size available for analysis is given as $n_{max}=\lambda t$.  
We assume that this is the sample size required to conduct $J^* k^*$ analyses.

Let us denote the number of treatment arms evaluated in $J^*$ analyses as $K_{eval}$. 
Then it can be determined as the number of $k$’s such that $j_{0k}+\psi_k\le J^*$. 
Note that, $K_{eval}$ is a random variable with a maximum value of $J^* k^*$ and a minimum of $[k_{init}+\frac{J^*}{J-1}]$ (assuming at least one experimental arm is evaluated for a maximum of J stages in the ongoing trial). 
In general, $k_{init} (\le k^*)$ is the number of arms the trial begins with.

Now, there are two possible cases to consider to evaluate $E_{J^*} [K_{eval}]$. 
\begin{itemize}
    \item Case I: Whenever an arm is dropped/ found to be effective, a new treatment arm replaces it, keeping a maximum $k^*$ arms in each stage fixed.
    \item Case II: A new arm can be added at any stage in the ongoing trial as long as the number of arms does not exceed $k^*$ at that stage keeping the platform more flexible.
\end{itemize}
	
While the first case allows us to analyse a maximum number of treatment arms, in reality, this might be harder to achieve as new treatment arms might not be readily available to be added in the trial. 
However, for this study we limit our investigation to case I only as for case II, the possibilities increases exponentially with $k^*$.
Further, Case I provides us with the maximum possible treatment arms that can be evaluated in $J^*$ analyses.
In this case, the minimum value of $K_{eval}$ can be found as $\left\lfloor \frac{J^*}{J} \right\rfloor*k^*$ with $k_{init}=k^*$, i.e. we run the platform trial at its maximum potential.
Here, $\left\lfloor \cdot \right\rfloor$ denotes the function defined as $\left\lfloor \frac{a}{b} \right\rfloor = \max\left\{n\in\mathbb{Z} : n \le \frac{a}{b}\right\}.$


There are multiple ways in which $K_{eval}=k_{eval}$ can occur. Let $\chi_{eval}$ denote the set of all possible combinations of treatment arms evaluated over $J^*$ stages such that exactly $k_{eval}$ treatment arms are evaluated, i.e., all possible combinations of $\{(j_{0k},\omega_k,\psi_k); k=1,2,\ldots,k_{eval}\}$ satisfying this condition.
For each element of $\chi_{eval}$, there is a unique set of upper $(u_{\chi_{eval}})$ and lower $(l_{\chi_{eval}})$ boundaries, containing a total of $\sum_{k=1}^{k_{eval}}\psi_k$ boundary values. These values are drawn from the set
\[
\{-\infty,f_1,\ldots,f_J,e_1,\ldots,e_J,\infty\}
\]
and occur in a specific order determined by the corresponding element of $\chi_{eval}$. Let $B_{eval}$ denote the set of all such boundary configurations. There is a one-to-one mapping between $\chi_{eval}$ and $B_{eval}$, such that each element of $\chi_{eval}$ uniquely determines the values and ordering of the corresponding lower $(l_{\chi_{eval}})$ and upper $(u_{\chi_{eval}})$ boundaries.

Therefore, the probability of $K_{eval}=k_{eval}$, denoted as $P[K_{eval}=k_{eval}]$ would be a sum of $n(\chi_{eval})$  ($ \sum_{k=1}^{k_{eval}}\psi_k $)-dimensional integral over a multivariate normal density function [here, n(.) denotes the cardinality of a set]. 
Each integral will obtain the upper and lower limits taking a single element from $B_{eval}$ (i.e. $(l_{\chi_{eval}})$  and $(u_{\chi_{eval}})$ ).

Then, the expected number of treatment arms being evaluated during the  $J^*$ analyses is given as
\begin{equation}
    E_{J^*} [K_{eval} ]=\sum_{k_{eval}=\left\lfloor \frac{J^*}{J} \right\rfloor k^*}^{J^* k^*} k_{eval}*P[K_{eval}=k_{eval} ]
\end{equation}

Note that, when we say $K_{eval}=k_{eval}$, we mean that $k_{eval}$ number of treatments have been decided upon, i.e. we know whether the treatments are effective or not.
We do not count treatment arms for whom the decision regarding treatment efficacy are pending (i.e. the decision at the interim analysis is to continue to the next stage).

Since the analytical distribution of $K_{eval}$ is extremely difficult to write down in a closed form, we assess the efficiency of a platform trial through simulation.

\subsection{Efficiency metric in presence of outcome delay}

Let us assume it takes $m$ months to observe the primary treatment outcome.
In this case, as recruitment continues during the $m$ months of awaiting outcome, the number of participants recruited to the platform at the time of a particular interim analysis becomes higher than the number who have outcome information available to contribute to that analysis.
Following previous notations, we term them as ‘pipeline’ participants and denote them as $n_{delay}$.
Therefore, to conduct the same $J^*$ analyses, we recruit more participants including the pipeline participants recruited during this $m$ delay period of awaiting treatment outcomes.

Under a uniform recruitment rate, pipeline counts at a given stage are assumed to be identical across treatment arms, irrespective of \(j_{0k}\). 
For example, with a recruitment rate of 10 participants per month, 3 treatment slots and a 6-month delay, each arm (both treatments and control) accumulates $15\ (=\frac{10}{3+1}*6)$ pipelines at Stage 1, accumulating a total of 45 pipelines for the treatment arms (we have not counted the pipelines for control arms as they get included in the following interim analysis).
However, pipeline counts may vary across stages, decreasing (e.g., from 15 to 10 each arm at stage 2, say) once a treatment arm approaches or reaches its target sample size.
Therefore, in summary, pipeline counts may vary across stages but are identical across treatment arms under a uniform recruitment rate.

The value of $n_{delay}$, defined as the total number of accumulated pipelines, is determined by the configuration of each individual element of $\chi_{eval}$, the set of combinations $\{(j_{0k},\omega_k,\psi_k); k=1,\ldots,k_{eval}\}$. 
Variations in these configurations induce different stopping patterns across stages, resulting in different numbers of accumulated pipelines and, consequently, different values of $n_{delay}$.
The higher the chances are for a treatment to stop early, the higher the chance for it to produce pipeline participants that do not contribute to the analyses of the trial.
These pipelines in return inflate the number of sample required to conduct the same $J^*$ analyses.
Therefore, $n_{delay}$ in a platform trial is a random variable.
Regardless of the delay length, $n_{delay}$ takes the value 0 when all treatments in $J^*$ analyses continue till $J$ stages and takes the maximum value when all the treatment arms in $J^*$ analyses are stopped in their first interim analysis.
However, it is difficult to write the complete distribution of $n_{delay}$ analytically as it is determined by the configuration of each individual element of $\chi_{eval}$, the set of combinations $\{(j_{0k},\omega_k,\psi_k); k=1,\ldots,k_{eval}\}$. 

We define the efficiency of a platform trial through the Expected Throughput $[ET]$ computed as the expected number of treatment arms evaluated per 1000 participants.
Mathematically, this can be expressed as:
\begin{equation}\label{Eq:efficiency}
    ET= E_{J^*}\left[\frac{K_{eval}}{n_{max}+n_{delay}}\right]*1000
\end{equation}

When the outcomes are immediately available, the number of pipelines fall to 0, reducing the denominator of the metric. 
Thus, under no delay, we can evaluate a higher number of treatment arms evaluated on average per 1000 trial participants.
Whereas, when delay increases, there is a higher chance that a trial would recruit more pipeline participants, leading to a higher denominator value, decreasing the number of treatment arms evaluated on average per 1000 trial participants.
Note that, the choice of 1000 is arbitrary and could be lower or higher without affecting any comparison of designs or relative effect of delay.

Therefore, the metric gives a comparable values of number of treatment arms being evaluated across different delay lengths irrespective of the time window specified.
For generalisablity, it is suggested to consider $J^*\ge2J$ to observe the set of all possible combinations of $\{(j_{0k},\omega_k,\psi_k); k=1,\ldots,k_{eval}\}$.
For the ease of computation, we consider $J^*$ to be multiples of $J$.
However, the results from the metric can be compared across different values of $J^*$ as the metric reports treatment arms evaluated per 1000 participants.

Since the analytical expression of the metric as well as $E[K_{eval}]$ is particularly difficult to obtain, we conduct simulations to observe the value of this metric.

\subsection{Designing the simulation}

In order to assess the efficiency of platform trials in presence of outcome delay, we use the metric defined in equation \ref{Eq:efficiency}. 
However, as mentioned in the above section, it is not possible to obtain the value of the metric analytically. 
Hence, we use simulation to obtain the value of $ET$ under different values of delay lengths.
The following steps were followed to conduct the simulation.

\begin{enumerate}
    \item Obtain a MAMS design with number of treatment arms equal to the number of slots $(k^*)$ in the design and number of stages being $J$  with marginal power control of $100(1-\beta)\%$ for $\delta^*$ and FWER of 5\%. 
    Determine the stage-wise stopping boundaries, sample size required $(n)$ for each stage then.

    \item Fix the total number of interim looks ($J^*$).[Note that, this is a multiple of the number of stages for each individual treatment arm $(J)$. 
    E.g. if we are considering a 3-stage design, then the number of interim looks can be 3,6,9 and so on. This allows complete evaluation of each treatment arm if they never stop early, giving a minimum number of treatment arms evaluated.]

    \item Specify a fraction of the pool of treatment arms evaluated which would be effective. 
    These appear in random order in each replicate. 
    This provides the treatment effects for the different treatment arms for simulating the results.
    For this example, this fraction can takes values 0, 0.25 and 1 respectively to indicate global null, a proportion of treatments being effective and the global alternative being true.

    \item Assume a uniform patient accrual rate ($\lambda$ participants/month) and a fixed delay length [say, $m=m_1$. This helps to compute a stage-wise pipeline patient estimate for each treatment arm. 

    \item Simulate data for each treatment arm and note the number of pipelines at each interim. 
    Note the decision for the treatment arm.
    Replace the no. of pipelines at that stage as 0 if the treatment is continued to the next stage as these patients will contribute to the analysis in the next stage(s). 

    \item These pipeline patients ($n_{delay}$) are added to the maximum sample size reflecting how many extra samples will be required to conduct same number of interim analyses. 
    Then the number of treatment arms evaluated is divided by the maximum sample size including the pipelines and multiplied by 1000. 
    It can provide an estimate of $ET$ the number of treatment arms evaluated per 1000 recruited patient in presence of delay for that replicate.

    \item Repeat steps [1-6] 10000 times to get the estimate of $ET$ for delay length $m=m_1$.

    \item Repeat steps [1-7] for different values of $m$.
\end{enumerate}

We have considered different values of delay lengths and different recruitment rates to observe the impact of delay under different situations.
$\lambda$ was considered to take values 2, 10 and 50 per month to indicate slow, moderate and fast recruitment respectively.

The exact values for the parameters for which the above simulation were conducted is listed in table \ref{tab:simulation parameters}.

The codes to reproduce the results can also be found here: \url{https://github.com/AritraMukherjee/Platform-trials-and-delay}
 
\begin{table}[htbp]
\centering
\caption{Parameters used for the simulation study}
\label{tab:simulation parameters}
\begin{tabular}{@{}ll@{}}
\toprule
\textbf{Design parameters for MAMS}                   &               \\ \midrule
Standardised treatment effect $(\delta^*)$   & 0.5           \\
Type I error                                 & 0.05          \\
Power                                        & 0.8           \\
Power control                                & Marginal      \\
Stopping boundary                            &               \\
\qquad Efficacy                              & OBF           \\
\qquad Futility                              & Fixed at 0    \\
No. of stages $(J)$                            & 3             \\
No. of arms $(k^*)$                              & 3             \\ \midrule
\textbf{Platform trial parameters}                    &               \\ \midrule
Proportion of effective   treatment arms     & 0, 0.25, 1    \\
Total no. of interim looks $(J^*)$           & 9             \\
No. of slots $(k^*)$                         & 3             \\
Delay $(m)$                                  & 0,3,6,9,12,18 \\
Recruitment rate $(\lambda)$(per month)      & 2, 10, 50      \\ \midrule
\textbf{Simulation parameters }                       &               \\ \midrule
No. of replicates                            & 10000         \\ \bottomrule
\end{tabular}
\end{table}

\section{Results}

In this section, we examine the effect of outcome delay on platform trial performance by presenting the distribution of $ET$ for delay lengths of 0, 3, 6, 9, 12, and 18 months. 
The results shown correspond to a recruitment rate of 10 participants per month, representing a moderate recruitment scenario.

\begin{figure}[htbp]
    \centering
    \includegraphics[width=0.75\linewidth]{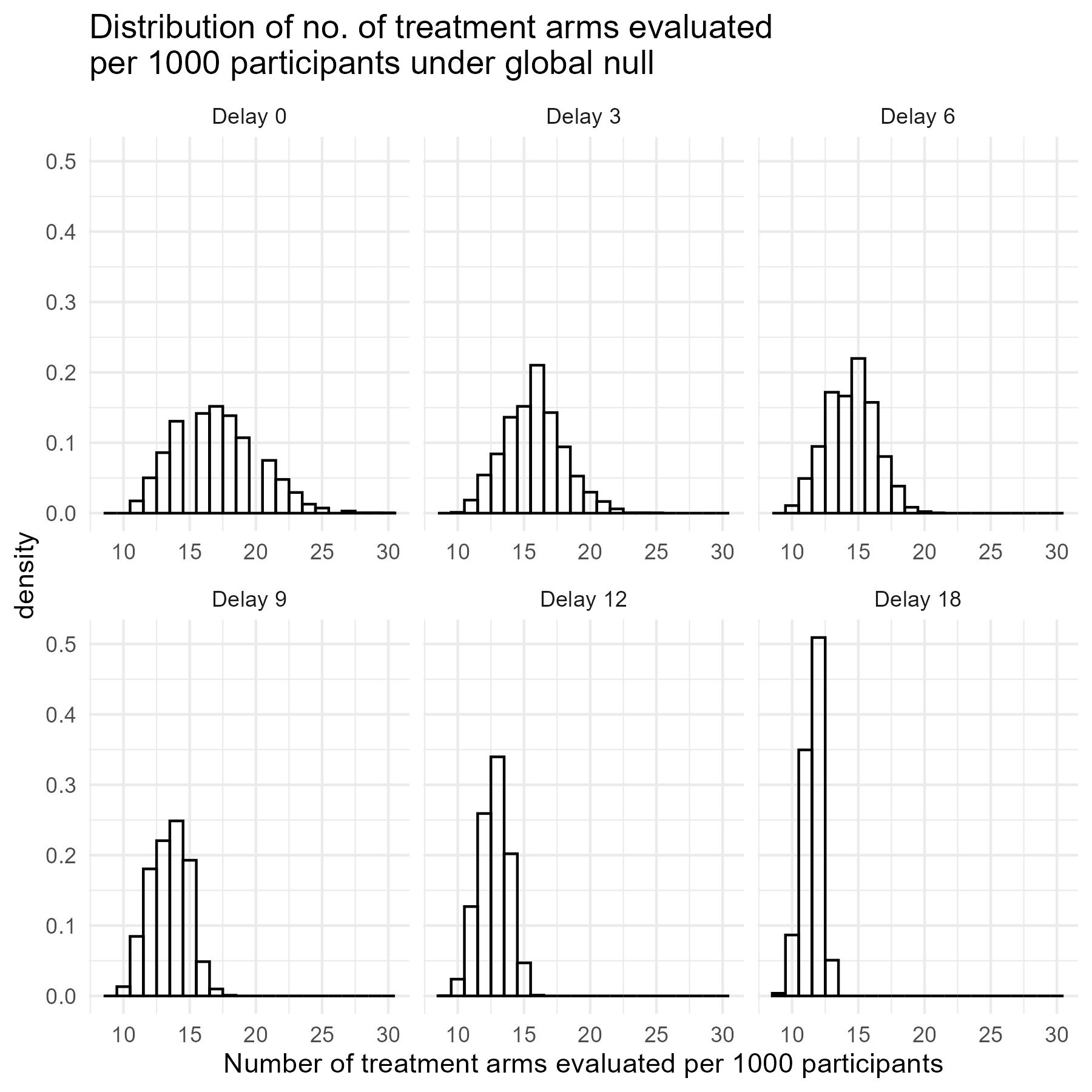}
    \caption{Distribution of the metric $ET=$ Treatment arms evaluated per 1000 participants for a platform trial with 3 treatment slots, a target treatment effect of $\delta^*=0.5$, $\alpha=0.05$, marginal power of 80\% and $J=3$ stages in a time window of 9 analyses across 10000 replicates. The impact of delay on the number of treatment evaluated here is determined assuming a recruitment rate of 10 under the assumption that the global null is true i.e. no treatment is effective.}
    \label{fig:1:Global_null_lambda_10}
\end{figure}

\begin{figure}[htbp]
    \centering
    \includegraphics[width=0.75\linewidth]{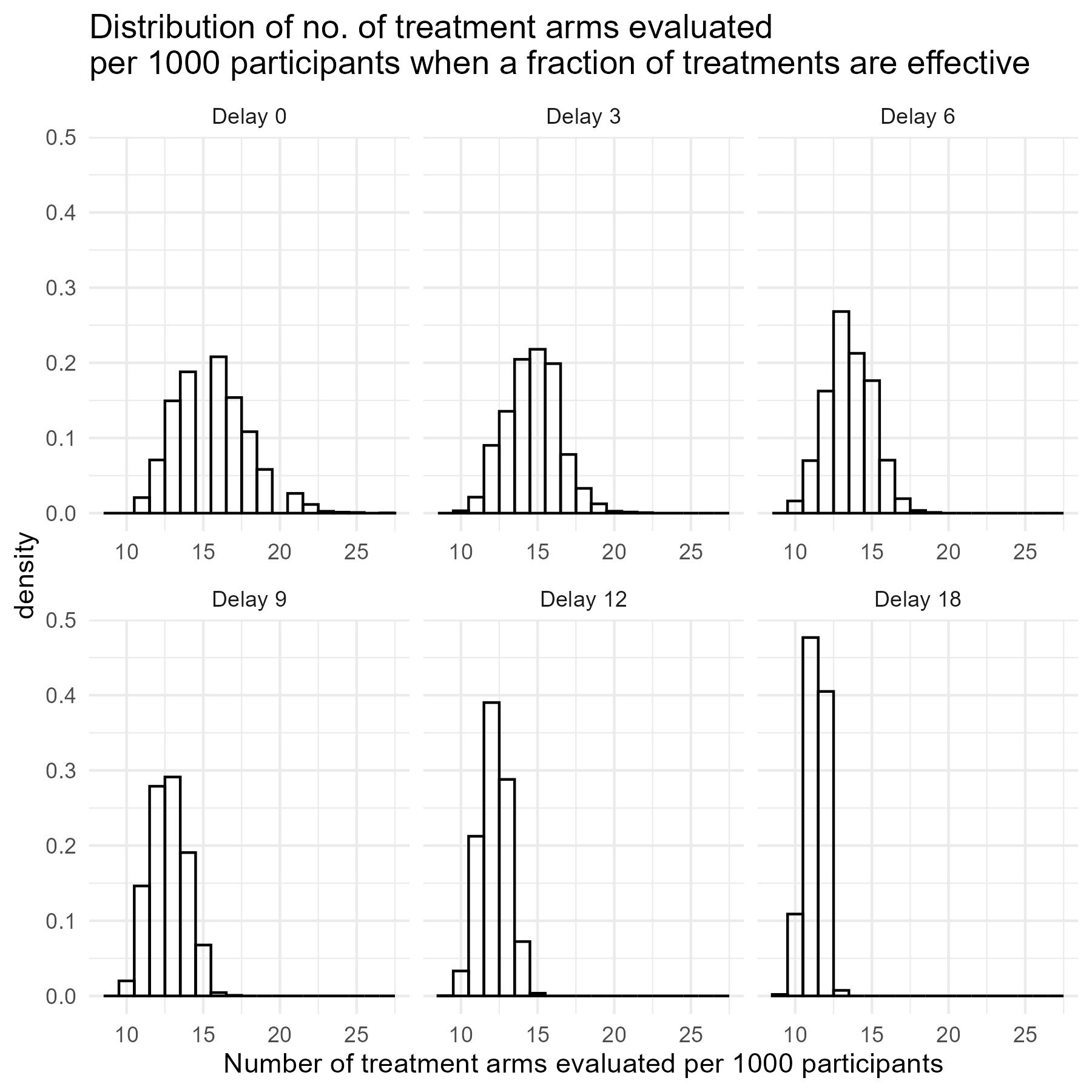}
    \caption{Distribution of the metric $ET=$ Treatment arms evaluated per 1000 participants for a platform trial with 3 treatment slots, a target treatment effect of $\delta^*=0.5$, $\alpha=0.05$, marginal power of 80\% and $J=3$ stages in a time window of 9 analyses across 10000 replicates. The impact of delay on the number of treatment evaluated here is determined assuming a recruitment rate of 10 under the assumption that a fraction of 0.25 of the total evaluated treatment arms are effective.}
    \label{fig:2:fract_effective_lambda_10}
\end{figure}

\begin{figure}[htbp]
    \centering
    \includegraphics[width=0.75\linewidth]{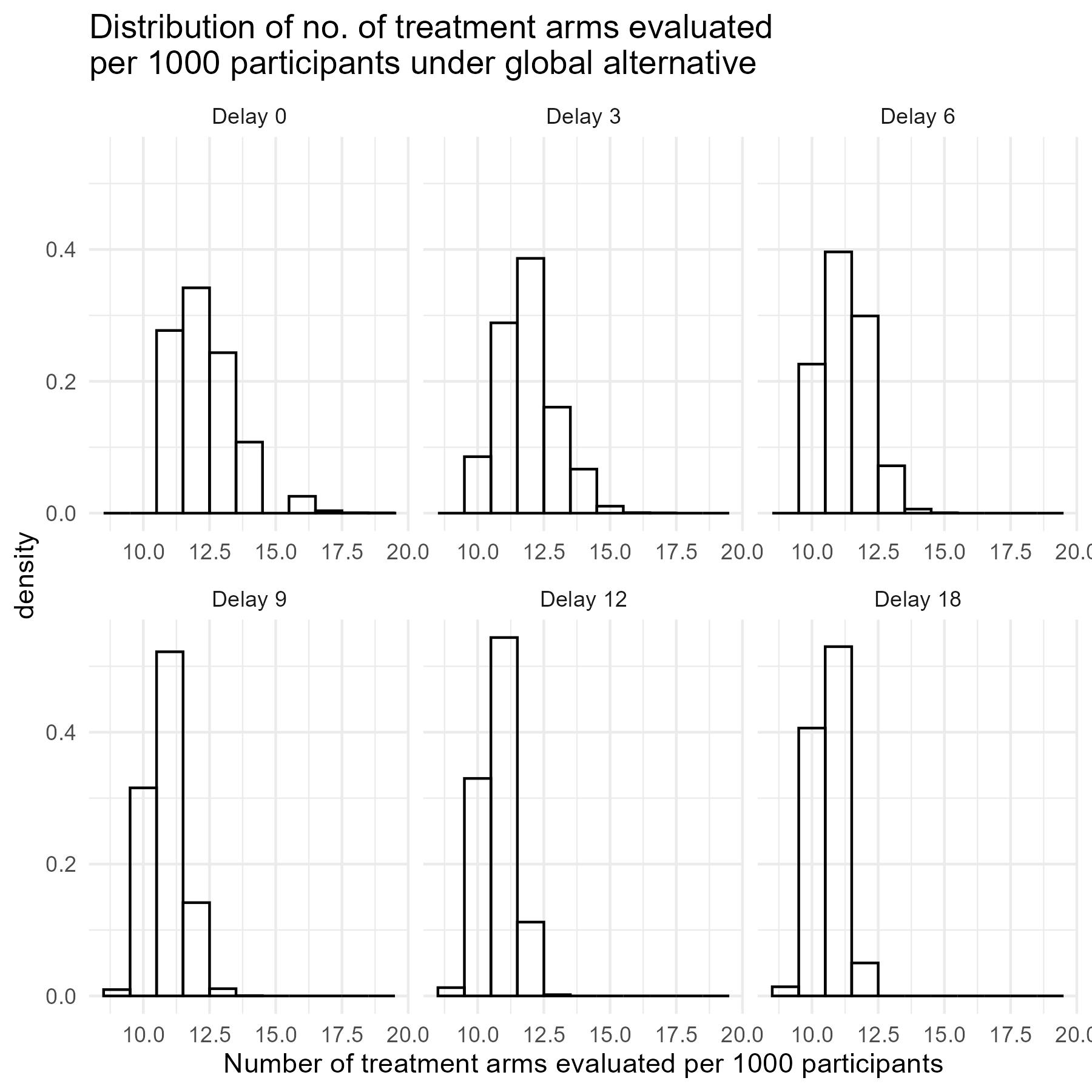}
    \caption{Distribution of the metric $ET=$ Treatment arms evaluated per 1000 participants for a platform trial with 3 treatment slots, a target treatment effect of $\delta^*=0.5$, $\alpha=0.05$, marginal power of 80\% and $J=3$ stages in a time window of 9 analyses across 10000 replicates. The impact of delay on the number of treatment evaluated here is determined assuming a recruitment rate of 10 under the assumption that the global alternative is true, i.e. all treatments are effective.}
    \label{fig:3:Global_alt_lambda_10}
\end{figure}

Figures \ref{fig:1:Global_null_lambda_10}, \ref{fig:2:fract_effective_lambda_10} and \ref{fig:3:Global_alt_lambda_10} display the distribution of $ET$ under the global null hypothesis, a scenario in which only a fraction ($1/4$) of the treatment arms are effective and the global alternative hypothesis respectively.

Corresponding figures for recruitment rates of 2 and 30 participants per month are provided in the Supplementary Material to illustrate the effects of slower and faster recruitment, respectively. 
For completeness, Table \ref{tab: Summary measures_efficiency metric} summarises the key descriptive measures of the efficiency metric across all three recruitment scenarios, providing a convenient basis for comparison.

For this platform trial, a total sample size of 828 participants is required to complete the nine interim analyses with three treatment slots under the assumption of no outcome delay. 
Here, each treatment arm at each stage requires approximately 23 participants or $n_{jk}=n=23$.
However, as the outcome delay increases, the required sample size rises rapidly because of the additional pipeline participants who accumulate between interim analyses.

Assuming a 10 per month recruitment rate, i.e. a recruitment rate of 2.5 participants per arm, the above platform design requires about 9 months to recruit for one stage and approximately 28 months to complete recruitment for an arm for all stages.
Therefore, this section provides results when the delay length is approximately $\frac{1}{10}, \frac{2}{10},\frac{3}{10},\frac{4}{10}^{th}$ and $\frac{2}{3}^{rd}$ of the total recruitment length for a single arm.

A consistent trend across all three figures is that the distribution of $ET$ gradually shifts to the left as the outcome delay increases, indicating a reduced likelihood of evaluating a larger number of treatment arms. 
This occurs because longer delays lead to the accumulation of more pipeline participants at each interim analysis. 
Consequently, a larger total sample size is required to complete the same number of interim analyses, resulting in a progressive decrease in the number of treatment arms evaluated per 1,000 participants.

The impact of outcome delay is more pronounced under the global null hypothesis because a fixed futility stopping boundary of 0 permits ineffective treatment arms to be discontinued early. 
Although early stopping for futility reduces the number of participants assigned to these arms, the outcome delay results in the accumulation of pipeline participants before the stopping decision can be implemented. 
Consequently, a larger number of pipeline participants accumulates across the trial.
As the proportion of effective treatment arms increases, early stopping for efficacy becomes less frequent because the O'Brien–Fleming (OBF)-type stopping boundary is intentionally conservative at early interim analyses. 
Consequently, fewer pipeline participants accumulate following early stopping decisions, reducing the impact of outcome delay. 
This pattern is clearly illustrated by comparing the results under the global null hypothesis, scenarios in which only a fraction of the treatment arms are effective, and the global alternative hypothesis, where all treatment arms are assumed to be effective.
Additional results (not shown) further confirm that the impact of outcome delay on the $ET$ metric diminishes as the proportion of effective treatment arms increases.

Results for other types of stopping boundaries are provided in detail in the Supplementary Materials. 
Table \ref{tab: Summary measures_stopping bounds} summarises the range of values observed for $ET$ across the different stopping boundaries for a recruitment rate of 10 participants per month.
The plots in the supplementary materials indicate that, under the global null, delay has a similar impact across all stopping boundaries, as the futility boundaries are fixed at zero. 
Under the global alternative, in the absence of delay, the O'Brien--Fleming (OBF) boundaries generally yield a narrower range of possible $ET$ values. 
This is primarily because treatment arms are more likely to continue until the final stage, thereby reducing the maximum possible $ET$. 
In contrast, Pocock and triangular boundaries facilitate earlier efficacy stopping, resulting in a more symmetric or evenly distributed $ET$ distribution under no delay. 
Consequently, the distribution of $ET$ under the OBF boundary appears to shift towards lower values more rapidly than those under the other boundary types.
In the presence of substantial delay (18 months or $\frac{2}{3}^{rd}$ of the recruitment length of a single arm), the differences between stopping boundaries become more pronounced. 
Under a Pocock efficacy boundary, more than half of the simulation replicates evaluated only 10 arms on average, whereas, under the other stopping boundaries, there was approximately a 40\% probability that more than 10 arms would be evaluated, with a maximum of 12 arms. 
This represents a reduction from a maximum of 21 arms evaluated per 100 participants under no delay. 
Similar patterns were observed when only a fraction of the administered treatment was effective.
Across all scenarios, the maximum value of $ET$ was capped at 12 under sufficiently large delays.

Table \ref{tab: Summary measures_efficiency metric} shows that the impact of outcome delay becomes increasingly pronounced as the recruitment rate increases. 
For example, even with a substantial outcome delay is 18 months, an average of 15 treatment arms can still be evaluated per 1,000 participants when the recruitment rate is 2 participants per month. 
In contrast, this average decreases to approximately 11 treatment arms for recruitment rates of 10 and 30 participants per month.

Similarly, under the higher recruitment rates, the maximum number of treatment arms evaluated per 1,000 participants does not exceed 13, whereas the corresponding maximum under the slower recruitment rate reaches as high as 23. 
These findings demonstrate that, depending on the recruitment rate, outcome delay has the potential to nearly halve the number of treatment arms that can be evaluated using the same number of participants.

An interesting observation from Table \ref{tab: Summary measures_efficiency metric} is that the minimum number of treatment arms evaluated per 1,000 participants for a given outcome delay differs across recruitment rates. 
For example, under the global null hypothesis, one might expect the minimum number of treatment arms evaluated per 1000 participants for a delay length of 9 months to be similar regardless of the recruitment rate, say, $10.58$, as observed for a slow recruitment rate. 
However, this is not seen in the simulation results.
This difference arises because the number of pipeline participants accumulated during the trial depends on the recruitment rate, which directly affects the denominator of the $ET$ metric. 
Faster recruitment leads to the accumulation of more pipeline participants before interim decisions can be implemented, thereby increasing the total number of participants required and reducing the value of $ET$. 
Consequently, the minimum efficiency values are slightly lower for higher recruitment rates, even when the outcome delay is the same. 
Nevertheless, these differences are relatively small in magnitude.

Additional results examining the impact of delayed outcomes across different numbers of treatment slots are provided in detail in the Supplementary Materials. 
Table \ref{tab: Summary measures_number of slots} summarises the mean, minimum, and maximum values of $ET$ under different delay lengths for an OBF stopping boundary, for reference.
The impact of delay appears to vary slightly across different numbers of treatment slots. 
This variation is primarily attributable to the assumption of a constant recruitment rate per month, reflecting the expectation that patient accrual in practice is unlikely to change according to the number of available treatment slots. 
Consequently, as the number of treatment slots increases, the recruitment rate per arm decreases, resulting in relatively slower recruitment within each arm. 
Thus, for a given delay length, the impact of the delay appears more pronounced as the number of treatment slots increases.
Nevertheless, the overall inference regarding the detrimental impact of increasing delay length remains consistent across all numbers of treatment slots.

We also conducted simulations for $J^* = 12$, and the corresponding results are presented in the Supplementary Material. 
The conclusions are consistent with those obtained for $J^* = 9$, further demonstrating that $ET$ provides a robust and informative summary measure for assessing and comparing the efficiency of platform trials, irrespective of the evaluation time window considered.

\begin{table}[]
\centering
\caption{Summary measures of the metric $ET=$Treatment arms evaluated per 1000 participants for a platform trial with 3 treatment slots, a target treatment effect of $\delta^*=0.5$, $\alpha=0.05$, marginal power of 80\% and $J=3$ stages in a time window of 9 analyses across 10000 replicates. The table records results for $ET$ for different types of stopping boundaries. The impact of delay on the number of treatment evaluated here is determined assuming a recruitment rate of 10 per month.}
\label{tab: Summary measures_stopping bounds}
\resizebox{\textwidth}{!}{%
\begin{tabular}{@{}lclllllllll@{}}
\toprule
\multicolumn{2}{c}{\textbf{Stopping boundary}} &
  \multicolumn{3}{c}{\textbf{Pocock}} &
  \multicolumn{3}{c}{\textbf{OBF}} &
  \multicolumn{3}{c}{\textbf{Triangular}} \\ \cmidrule(l){3-11}
\multicolumn{2}{c}{$ET$} &
  \textit{Mean} &
  \textit{Min} &
  \textit{Max} &
  \textit{Mean} &
  \textit{Min} &
  \textit{Max} &
  \textit{Mean} &
  \textit{Min} &
  \textit{Max} \\ \midrule
  \textbf{Case} & \textbf{Delay}   &       &       &       &       &       &       &       &       &  \\ \bottomrule
Global null        & 0  & 14.64 & 9.26 & 24.69 & 17.07 & 10.87 & 30.19 & 15.72 & 10.00 & 27.78 \\
                   & 3  & 13.56 & 8.98 & 20.97 & 15.63 & 10.40 & 24.62 & 14.49 & 9.60  & 22.99 \\
                   & 6  & 12.65 & 8.72 & 18.22 & 14.44 & 9.97  & 20.78 & 13.46 & 9.23  & 19.61 \\
                   & 9  & 11.86 & 8.47 & 16.11 & 13.43 & 9.57  & 17.98 & 12.58 & 8.89  & 17.09 \\
                   & 12 & 11.22 & 8.29 & 14.47 & 12.71 & 9.47  & 15.98 & 11.90 & 8.78  & 15.24 \\
                   & 18 & 10.26 & 8.06 & 12.07 & 11.52 & 9.33  & 13.10 & 10.85 & 8.65  & 12.59 \\ \hline
Fraction effective & 0  & 14.26 & 9.26 & 23.66 & 15.65 & 10.87 & 26.57 & 15.11 & 10.00 & 25.56 \\
                   & 3  & 13.25 & 8.98 & 20.23 & 14.52 & 10.40 & 22.32 & 13.99 & 9.68  & 21.60 \\
                   & 6  & 12.39 & 8.72 & 17.67 & 13.55 & 9.97  & 19.25 & 13.04 & 9.38  & 18.70 \\
                   & 9  & 11.64 & 8.47 & 15.68 & 12.71 & 9.57  & 17.03 & 12.22 & 9.09  & 16.49 \\
                   & 12 & 11.04 & 8.29 & 14.15 & 12.19 & 9.47  & 15.35 & 11.63 & 8.91  & 14.84 \\
                   & 18 & 10.18 & 8.06 & 11.94 & 11.29 & 9.33  & 12.82 & 10.72 & 8.65  & 12.43 \\ \hline
Global alternative & 0  & 13.13 & 9.26 & 21.60 & 12.41 & 10.87 & 19.32 & 13.34 & 10.00 & 21.11 \\
                   & 3  & 12.27 & 8.85 & 18.72 & 11.83 & 10.31 & 17.01 & 12.49 & 9.52  & 18.36 \\
                   & 6  & 11.53 & 8.47 & 16.51 & 11.30 & 9.80  & 15.19 & 11.74 & 9.09  & 16.24 \\
                   & 9  & 10.87 & 8.13 & 14.77 & 10.83 & 9.35  & 13.95 & 11.09 & 8.70  & 14.56 \\
                   & 12 & 10.40 & 7.94 & 13.44 & 10.73 & 9.32  & 13.27 & 10.73 & 8.57  & 13.48 \\
                   & 18 & 9.84  & 7.94 & 11.57 & 10.59 & 9.32  & 12.26 & 10.29 & 8.57  & 11.95 \\ \hline
\end{tabular}%
}
\end{table}

\begin{table}[]
\centering
\caption{Summary measures of the metric $ET=$ Treatment arms evaluated per 1000 participants for a platform trial with 3 treatment slots, a target treatment effect of $\delta^*=0.5$, $\alpha=0.05$, marginal power of 80\% and $J=3$ stages in a time window of 9 analyses across 10000 replicates. The impact of delay on the number of treatment evaluated here is determined assuming different recruitment rates.}
\label{tab: Summary measures_efficiency metric}
\resizebox{\textwidth}{!}{%
\begin{tabular}{@{}lclllllllll@{}}
\toprule
\multicolumn{2}{c}{\textbf{Recruitment rate}} &
  \multicolumn{3}{c}{\textbf{2 per month}} &
  \multicolumn{3}{c}{\textbf{10 per month}} &
  \multicolumn{3}{c}{\textbf{30 per month}} \\ \cmidrule(l){3-11}
\multicolumn{2}{c}{$ET$} &
  \textit{Mean} &
  \textit{Min} &
  \textit{Max} &
  \textit{Mean} &
  \textit{Min} &
  \textit{Max} &
  \textit{Mean} &
  \textit{Min} &
  \textit{Max} \\ \midrule
  \textbf{Case} & \textbf{Delay}   &       &       &       &       &       &       &       &       &  \\ \bottomrule
Global null        & 0  & 17.03 & 10.87 & 28.99 & 17.07 & 10.87 & 30.19 & 17.05 & 10.87 & 28.99 \\
                   & 3  & 16.72 & 10.77 & 27.83 & 15.63 & 10.40  & 24.62 & 13.43 & 9.80  & 17.84 \\
                   & 6  & 16.42 & 10.68 & 26.76 & 14.44 & 9.97  & 20.78 & 11.52 & 9.34  & 13.04 \\
                   & 9  & 16.14 & 10.58 & 25.76 & 13.43 & 9.57  & 17.98 & 11.45 & 9.32  & 12.88 \\
                   & 12 & 15.86 & 10.49 & 24.84 & 12.71 & 9.47  & 15.98 & 11.45 & 9.32  & 12.88 \\
                   & 18 & 15.35 & 10.31 & 23.19 & 11.52 & 9.33  & 13.1  & 11.45 & 9.32  & 12.88 \\ \hline
Fraction effective & 0  & 15.60 & 10.87 & 30.19 & 15.65 & 10.87 & 26.57 & 15.68 & 10.87 & 25.36 \\
                   & 3  & 15.36 & 10.77 & 28.89 & 14.52 & 10.40  & 22.32 & 12.73 & 9.57  & 16.73 \\
                   & 6  & 15.13 & 10.68 & 27.69 & 13.55 & 9.97  & 19.25 & 11.29 & 9.33  & 12.73 \\
                   & 9  & 14.91 & 10.58 & 26.58 & 12.71 & 9.57  & 17.03 & 11.24 & 9.32  & 12.60 \\
                   & 12 & 14.69 & 10.49 & 25.56 & 12.19 & 9.47  & 15.35 & 11.24 & 9.32  & 12.60 \\
                   & 18 & 14.28 & 10.31 & 23.74 & 11.29 & 9.33  & 12.82 & 11.24 & 9.32  & 12.60 \\ \hline
Global alternative & 0  & 12.42 & 10.87 & 18.12 & 12.41 & 10.87 & 19.32 & 12.41 & 10.87 & 18.12 \\
                   & 3  & 12.30 & 10.75 & 17.73 & 11.83 & 10.31 & 17.01 & 10.82 & 9.35  & 13.30 \\
                   & 6  & 12.18 & 10.64 & 17.36 & 11.30  & 9.80   & 15.19 & 10.58 & 9.32  & 12.12 \\
                   & 9  & 12.06 & 10.53 & 17.01 & 10.83 & 9.35  & 13.95 & 10.58 & 9.32  & 12.08 \\
                   & 12 & 11.95 & 10.42 & 16.67 & 10.73 & 9.32  & 13.27 & 10.58 & 9.32  & 12.08 \\
                   & 18 & 11.73 & 10.20 & 16.03 & 10.59 & 9.32  & 12.26 & 10.58 & 9.32  & 12.08 \\  \bottomrule
\end{tabular}%
}
\end{table}

\begin{table}[]
\centering
\caption{Summary measures of the metric $ET=$ Treatment arms evaluated per 1000 participants for a platform trial with different number of treatment slots $(k^*)$, a target treatment effect of $\delta^*=0.5$, $\alpha=0.05$, marginal power of 80\% and $J=3$ stages in a time window of 9 analyses across 10000 replicates. The underlying design uses an OBF type stopping boundary. The impact of delay on the number of treatment evaluated here is determined assuming a recruitment rate of 10 per month.}
\label{tab: Summary measures_number of slots}
\resizebox{\textwidth}{!}{%
\begin{tabular}{@{}lclllllllll@{}}
\toprule
\multicolumn{2}{c}{$k^*$} &
  \multicolumn{3}{c}{\textbf{2}} &
  \multicolumn{3}{c}{\textbf{3}} &
  \multicolumn{3}{c}{\textbf{4}} \\ \cmidrule(l){3-11}
\multicolumn{2}{c}{$ET$} &
  \textit{Mean} &
  \textit{Min} &
  \textit{Max} &
  \textit{Mean} &
  \textit{Min} &
  \textit{Max} &
  \textit{Mean} &
  \textit{Min} &
  \textit{Max} \\ \midrule
  \textbf{Case} & \textbf{Delay}   &       &       &       &       &       &       &       &       &  \\ \bottomrule
Global null        & 0  & 16.56 & 10.58 & 31.75 & 17.07 & 10.87 & 30.19 & 16.74 & 10.67 & 28.44 \\
                   & 3  & 14.79 & 9.88  & 24.10 & 15.63 & 10.40 & 24.62 & 15.62 & 10.28 & 24.41 \\
                   & 6  & 13.40 & 9.27  & 19.42 & 14.44 & 9.97  & 20.78 & 14.65 & 9.93  & 21.38 \\
                   & 9  & 12.45 & 9.22  & 16.26 & 13.43 & 9.57  & 17.98 & 13.81 & 9.59  & 19.01 \\
                   & 12 & 11.66 & 9.22  & 13.99 & 12.71 & 9.47  & 15.98 & 13.07 & 9.28  & 17.12 \\
                   & 18 & 11.52 & 9.22  & 13.61 & 11.52 & 9.33  & 13.10 & 12.01 & 9.15  & 14.42 \\ \hline
Fraction effective & 0  & 15.41 & 10.58 & 28.22 & 15.65 & 10.87 & 26.57 & 15.30 & 10.67 & 25.78 \\
                   & 3  & 13.94 & 9.88  & 22.01 & 14.52 & 10.40 & 22.32 & 14.43 & 10.34 & 22.53 \\
                   & 6  & 12.75 & 9.27  & 18.04 & 13.55 & 9.97  & 19.25 & 13.66 & 10.03 & 20.01 \\
                   & 9  & 12.03 & 9.22  & 15.55 & 12.71 & 9.57  & 17.03 & 12.97 & 9.73  & 18.00 \\
                   & 12 & 11.42 & 9.22  & 13.69 & 12.19 & 9.47  & 15.35 & 12.35 & 9.46  & 16.36 \\
                   & 18 & 11.31 & 9.22  & 13.37 & 11.29 & 9.33  & 12.82 & 11.58 & 9.41  & 13.90 \\ \hline
Global alternative & 0  & 12.42 & 10.58 & 19.40 & 12.41 & 10.87 & 19.32 & 12.04 & 10.67 & 17.78 \\
                   & 3  & 11.59 & 9.88  & 16.49 & 11.83 & 10.31 & 17.01 & 11.61 & 10.23 & 16.30 \\
                   & 6  & 10.89 & 9.27  & 14.34 & 11.30 & 9.80  & 15.19 & 11.21 & 9.83  & 15.05 \\
                   & 9  & 10.71 & 9.22  & 13.38 & 10.83 & 9.35  & 13.95 & 10.84 & 9.46  & 13.98 \\
                   & 12 & 10.58 & 9.22  & 12.61 & 10.73 & 9.32  & 13.27 & 10.50 & 9.11  & 13.05 \\
                   & 18 & 10.56 & 9.22  & 12.47 & 10.59 & 9.32  & 12.26 & 10.36 & 9.06  & 12.19 \\ \hline
\end{tabular}%
}
\end{table}

\section{Discussion}

Previous studies on the impact of outcome delay on adaptive designs have been indicating the possible considerable efficiency losses in presence of large outcome delay.
Platform trials with its many advantages is not immune to this impact of a delayed outcome.
If it takes a long time to observe the primary treatment outcome, the trial recruits a good number of pipeline patients who do not benefit from the adaptation of stopping the trial early in a MAMS-platform.
In this work, we aim to quantify this loss in efficiency through proposing a novel metric: $ET$.
We proposed $ET$ as a simple and interpretable metric for quantifying the operational efficiency of platform trials by measuring the number of treatment arms evaluated per 1,000 participants. 
Using extensive simulation studies, we investigated how this metric is affected by outcome delay, recruitment rate, and the underlying proportion of effective treatment arms.

Across all simulation scenarios, outcome delay consistently reduced platform efficiency. 
As the delay between participant recruitment and outcome availability increased, a greater number of pipeline participants accumulated before interim decisions could be implemented. 
Consequently, additional participants were enrolled into treatment arms that would otherwise have been stopped, increasing the overall sample size required to complete the planned sequence of interim analyses. 
This translated into a reduction in the number of treatment arms that could be evaluated for a fixed number of participants, demonstrating the operational cost of delayed outcomes in adaptive platform trials.

The magnitude of this reduction was strongly influenced by the recruitment rate. 
Under faster recruitment, participants accumulated more rapidly during the outcome delay period, leading to substantially larger pipeline populations. 
As a result, the efficiency loss associated with delayed outcomes was considerably greater than under slower recruitment. 
In some scenarios, increasing the outcome delay from 0 to 18 months nearly halved the number of treatment arms that could be evaluated per 1,000 participants under higher recruitment rates. 
It was observed that a delay length of more than $2/3^{rd}$ of the total recruitment length for a single arm results in considerably higher loss of efficiency in terms of reducing the $ET$.
This is consistent with our earlier findings across other types of adaptive designs \cite{Mukherjee2022Cancer,Mukherjee2025GSD,Mukherjee2025MAMS}.
However, a platform trial appears to be more robust to efficiency loss due to delay compared to a MAMS or GSD.
This can be attributed to generation of a lower number of pipelines in the control arm, compared to an equivalent MAMS or a GSD.
These findings highlight the important interaction between recruitment speed and outcome delay and suggest that recruitment rate should be considered alongside endpoint timing when designing adaptive platform trials.

The impact of outcome delay also depended on the underlying treatment effects. 
Under the global null hypothesis, delayed outcomes produced the largest reduction in efficiency. 
This is because the futility stopping boundary allowed ineffective treatment arms to be discontinued relatively early, but outcome delays postponed these decisions, resulting in the accumulation of additional pipeline participants. 
In contrast, when a greater proportion of treatment arms were effective, the reduction in efficiency became progressively smaller. 
The conservative O'Brien--Fleming efficacy boundaries made early stopping for efficacy relatively uncommon, limiting the number of participants enrolled after a treatment had effectively crossed the stopping boundary. Consequently, fewer pipeline participants accumulated, reducing the adverse impact of delayed outcomes. 
This pattern was consistently observed across scenarios ranging from the global null to the global alternative.
Further, the choice of stopping boundaries can influence of the magnitude of the delay impact.


An important finding was that the qualitative conclusions remained unchanged when the evaluation window was extended from $J^*=9$ to $J^*=12$. 
This robustness suggests that $ET$ is not sensitive to the specific time horizon chosen for evaluation and can therefore serve as a general summary measure for comparing the operational efficiency of alternative platform trial designs.

The proposed metric complements conventional operating characteristics such as power, type I error, sample size, and trial duration by explicitly quantifying how efficiently participant resources are converted into treatment evaluations. 
As platform trials continue to become more common, particularly in settings where multiple interventions are assessed sequentially or concurrently, such an operational measure may provide useful additional insight when comparing competing design options.

Several limitations should be acknowledged. 
The present study focused on fixed recruitment rates and specific delay lengths under a particular adaptive platform design. 
The underlying assumption of a uniform accrual of participants can be thought to be the `steady state' of the platform trial once it has opened all its sites.
Different recruitment patterns, staggered site activation (leading to a `Mixed recruitment' pattern \cite{Mukherjee2025GSD}), time-varying recruitment, alternative interim analysis schedules, or different stopping rules may influence the absolute values of $ET$. 
Nevertheless, the underlying mechanism driving efficiency loss, i.e. the accumulation of pipeline participants during delayed outcome assessment, is common to many adaptive platform trials, suggesting that the qualitative findings are likely to be broadly applicable. 
Future work could extend the proposed framework to more complex platform designs incorporating adaptive randomisation schemes and settings with random outcome delays.

Overall, the results demonstrate that outcome delay can substantially reduce the operational efficiency of platform trials, particularly when recruitment is rapid. 
The proposed $ET$ metric provides an intuitive and robust framework for quantifying this loss and comparing alternative platform trial designs. 
Incorporating such efficiency measures alongside traditional operating characteristics may enable more informed design choices that better balance statistical performance with participant resource utilisation.

\section*{Acknowledgments}

JMSW and AM are funded by a NIHR Research Professorship (NIHR301614). 

\printbibliography

\end{document}